\documentclass[aip,jcp,amscd,twocolumn]{revtex4}
\usepackage{amssymb}
\usepackage{amsmath}

\begin{document}

\title{On the equivalence of LIST and DIIS methods for 
  convergence acceleration}
\author{Alejandro J. Garza}
\affiliation{Department of Chemistry, Rice University, Houston, Texas, 77251-1892, USA}
\author{Gustavo E. Scuseria}
\affiliation{Department of Chemistry and Department of Physics and Astronomy, Rice
University, Houston, Texas, 77251-1892, USA.}
\date{\today}

\begin{abstract}
Self-consistent field extrapolation methods play a pivotal role 
in quantum chemistry and electronic structure theory. 
We here demonstrate the mathematical equivalence between the 
recently proposed family of LIST methods
[J. Chem. Phys. \textbf{134}, 241103 (2011);
J. Chem. Theory Comput. \textbf{7}, 3045 (2011)]
with Pulay's DIIS [Chem. Phys. Lett. \textbf{73}, 393 (1980)].
Our results also explain the differences in performance 
among the various LIST methods. 
\end{abstract}

\pacs{0000-1111-222}
\maketitle

\label{I1}


\emph{\textbf{Introduction.}}
The importance of self-consistent field (SCF) extrapolation in 
Hartree--Fock and Kohn--Sham calculations cannot be overstated: 
without efficient techniques like DIIS~\cite{Pulay1,Pulay2}, 
much of the current body 
of work in computational quantum chemistry and related fields would 
not be practical. These techniques have permeated the quantum 
chemistry field and are ubiquitous in most (if not all) software packages.
DIIS has found its way to many other applications like extrapolation 
of coupled cluster solutions~\cite{Scuseria1986} and 
convergence acceleration in other computational science areas;
\emph{e.g.}, the popular GMRES method of Saad
and Schultz~\cite{Saad1986} is equivalent to DIIS
for linear problems~\cite{Schneider2012}. 

Since Pulay's seminal paper~\cite{Pulay1}, 
there has been an enormous amount of work to refine, 
improve, and complement 
DIIS~\cite{Cances2000,EDIIS,ADIIS,ARH,LeBris}
but the core of the technique remains the same, as originally formulated.
More recently, there have been claims in the 
literature~\cite{LISTi,LISTb}
that a family of recently proposed LIST methods supersedes the 
performance of DIIS. 
We here prove that the LIST methods are mathematically equivalent 
to DIIS, and as such, they can hardly be better than DIIS, 
as shown numerically in our previous work~\cite{Garza2012}. 
The results presented here should settle this controversy for good.


\emph{\textbf{DIIS.}}
We begin by writing DIIS in its general formulation 
as first outlined by Pulay~\cite{Pulay1}.
An improved density matrix $\tilde{D}_{k}$ 
is constructed from a linear 
combination of previously iterated density matrices 
\begin{equation}
  \tilde{D}_{k}=\sum\limits_{j=1}^{k}c_{j} D_{j}.
  \label{eq:expansion}
\end{equation}
The coefficients in the above equation are determined as 
\begin{equation}
  \{c_{j}\}=\text{arg min}\left\{ \left\Vert
  \sum\limits_{j=1}^{k}c_{j} {e}_j \right\Vert
  ,\sum\limits_{j=1}^{k}c_{j}=1\right\}
  \label{eq:DIIS}
\end{equation}%
where ${e}_j$ is an approximate error function associated with $D_j$,
and $|| \cdot ||$ denotes a suitable norm. 
That is, DIIS consists of creating a $\tilde{D}_{k}$ which
minimizes (in the least squares sense)  ${e}_j$ constrained 
to  $\text{Tr}(\tilde{D}_{k}) = N$, where $N$ is the number of 
electrons.

There are three LIST methods: LISTi, LISTd,~\cite{LISTi} and 
LISTb~\cite{LISTb}. We first show how LISTi is equivalent to 
the general DIIS described above.


\emph{\textbf{LISTi.}}
Let
$D_i^\text{in}$ and $D_i^\text{out}$ denote input and output 
density matrices from the diagonalization of the Fock matrix
$F_i^\text{in}$.
The equations to solve for the LISTi coefficients $\{ c_j \}$
can be written as (Ref.~\cite{LISTi}, Eq. 13)
\begin{equation}
  \sum_j c_j g_{ij} = 
  \sum_j c_j \text{Tr} ((D_j^\text{out} - D_j^\text{in})
  (F_i^\text{out} - F_i^\text{in})) = 0, 
  \label{eq:list}
\end{equation}
$  \forall i$,
restricted to $\sum_j c_j = 1$.
Assuming summation over repeated indices, 
Eq.~\ref{eq:list} is equivalent to 
$\left\vert c_j g_{ij} \right\vert = 0 $, 
$  \forall i$,
where $\left\vert \cdot \right\vert$ denotes absolute value. 
It is therefore also equivalent to 
$\text{max}
  \left\{ \left\vert c_j g_{ij} \right\vert ,
  i \in I \right\} = 0$
with $I = \{ 1, 2, ..., k \}$, where $k$ is the dimension of the 
iterative subspace.
Thus, the set $\{ c_j \}$  can be expressed as 
\begin{equation}
  \{ c_j \} = \text{arg min} \left\{ \text{max}
  \left\{ \left\vert \sum_j c_j g_{ij} \right\vert , i \in I
  \right\} , \sum_j c_j = 1 \right\}  .
  \label{eq:eq4}
\end{equation}
Noting that the infinity norm of a vector $\vec{v}$ of length 
$k$ is defined as $ || \vec{v} ||_\infty =
\text{max} \left\{ |v _1 |, ..., | v_k | \right\}$, 
Eq.~\ref{eq:eq4} becomes
\begin{equation}
  \{ c_j \} = \text{arg min}  
  \left\{ \left\Vert \sum_j c_j g_{ij} \right\Vert_\infty
  , \sum_j c_j = 1 \right\} , 
  \label{eq:eq5}
\end{equation}
Furthermore, if we let $x$ be an index such that 
\begin{equation}
  \left\vert \sum_j c_j g_{xj} \right\vert = 
  \left\Vert \sum_j c_j g_{ij} \right\Vert_\infty, 
\end{equation}
then we see that LISTi can be formulated exactly in 
the form of Eq.~\ref{eq:DIIS} with 
$e_j = g_{xj}$ and $|| \cdot || = || \cdot ||_\infty$. 

In deriving the above equivalence, we have not made use 
of the explicit form of the matrix $g_{ij}$. 
This has the corollary that any method that consists 
of solving a linear system of the form of 
Eq.~\ref{eq:list} (\emph{i.e.}, $\sum_j c_j b_{ij} = 0, \forall i, 
\sum c_j =1$) corresponds to the general DIIS 
minimization problem of Eq.~\ref{eq:DIIS}.
As we discuss below, LISTd and LISTb
can also be written in this form. 

%
%


\emph{\textbf{LISTd.}}
We proceed now to outline the relationship between DIIS 
and LISTd. These results also clarify 
the reason for the poor convergence acceleration of 
LISTd~\cite{LISTi,LISTb,Garza2012}.
In brief,
the better convergence properties of LISTb and LISTi
as compared to LISTd can be understood in terms of the properties
of the error function being minimized;
an SCF solution minimizes the error function in LISTb and LISTi, 
but not necessarily in LISTd.

The equations to be solved for the LISTd coefficients $\{ c_j \}$ 
are (Ref.~\cite{LISTi}, Eq. 10)
\begin{equation}
  \sum_j c_j a_{ij} = 
  \sum_j c_j \left[ E_i - E + \text{Tr}((D_j^\text{out} - D_i^\text{out}) 
    \Delta F_i) \right] = 0 , 
  \label{eq:listd}
\end{equation}
$\forall i$, with $\sum_j c_j = 1$,  
where $\Delta F_i = F_i^\text{out} - F_i^\text{in}$, and
$E$ is the current best estimate for the energy.
From the results of the previous section, it is already clear 
that Eq.~\ref{eq:listd} fits in the general framework 
of Eq.~\ref{eq:DIIS}.
We can also express $\{ c_j \}$ as 
\begin{equation}
  \{ c_j \} = \text{arg min} \left\{ \text{max}
  \left\{ \left\vert \sum_j c_j a_{ij} \right\vert , i \in I
  \right\} , \sum_j c_j = 1 \right\} ,
  \label{eq:minimax1}
\end{equation}
where we could also have used the infinity norm as in 
Eq.~\ref{eq:eq5}.
Thus, we can view LISTd as minimizing the approximate error function 
$e(\mathbf{c}) = \text{max}
\left\{ \left\vert \sum_j c_j a_{ij} \right\vert , i \in I
\right\}$ or, equivalently in terms of the extrapolated $\tilde{D}$, 
$e(\tilde{D}) = \text{max}
\left\{ \left\vert a_{i} (\tilde{D}) \right\vert , i \in I  
\right\}$ where 
\begin{equation}
  a_{i} (\tilde{D}) = 
  E_i - E + \text{Tr}((\tilde{D} - D_i^\text{out}) 
  \Delta F_i).
\end{equation}
Let now $E$ be the energy for the converged density matrix $D$. 
It is easy to see that $e(D) \neq 0$ in general because 
the matrices in the iterative subspace $D_i$ do not need 
(and are, in fact, not expected) to be converged.
Hence, by formulating LISTd as an error minimization problem, 
we see that the choice for the approximate error function is 
a poor one, since a converged density matrix can have nonzero
error---and therefore, it does not 
satisfy the LISTd equations.
This explains the unsatisfactory convergence acceleration of LISTd.


\emph{\textbf{LISTb.}}
The LISTb equations are given by the transpose of the LISTd 
matrix (Ref.~\cite{LISTb}, Eq. 14)
\begin{equation}
  \sum_j c_j a_{ij}' = 
  \sum_j c_j \left[ E_j - E + \text{Tr}((D_i^\text{out} - D_j^\text{out}) 
    \Delta F_j) \right] = 0, 
  \label{eq:listb}
\end{equation}
$\forall i$, with $\sum_j c_j = 1$.
Again, the results on the LISTi section imply that 
Eq.~\ref{eq:listb} corresponds to the general 
DIIS minimization problem of Eq.~\ref{eq:DIIS}.
Analogously to Eq. \ref{eq:minimax1}, we can write
\begin{equation}
  \{ c_j \} = \text{arg min} \left\{ \text{max}
  \left\{ \left\vert \sum_j c_j a_{ij}' \right\vert , i \in I
  \right\} , \sum_j c_j = 1 \right\}  ,
  \label{eq:minimax2}
\end{equation}
and the error function being minimized can be written as 
$e(\{ D_j \}, \mathbf{c} ) = \text{max}
\left\{ \left\vert a_{i} (\{ D_j \}, \mathbf{c}  ) \right\vert , i \in I  
\right\}$, with 
\begin{align}
  a_{i} (\{ D_j \}, \mathbf{c}  ) =
  \sum_j c_j \left[ E_j - E + \text{Tr}((D_i^\text{out} - D_j^\text{out}) 
    \Delta F_j) \right].
\end{align}
Because of the transposition, the error function can no longer 
be written in terms of the extrapolated matrix $\tilde{D}$. 
However, we can consider the error function for 
a set of matrices $\{ D_j \}$ for which $D_k = D$ 
is a converged density matrix.
When $E$ is the converged energy, then it is straightforward to 
see that, for a set of coefficients $\mathbf{c}$ with 
$c_j = 1$ if $j=k$ and $c_j = 0$ otherwise,  
$e(\{ D_j \}, \mathbf{c} ) = e(D,1) = 0$. 
Thus, a converged density matrix minimizes the LISTb error, 
satisfying Eq. \ref{eq:listb}.
In consequence, LISTb can correctly select $D_k = D$  
as an SCF solution from the set $\{ D_j \}$, whereas 
LISTd (in general) does not do this. 
The improved convergence acceleration of LISTb as compared to 
LISTd can therefore be attributed to the minimization of a 
more suitable error function (which is in fact
the idea behind the DIIS procedure).
The same applies to LISTi; 
since $D_j^\text{out} - D_j^\text{in}$ must be the zero matrix 
for a converged density, it is straightforward to 
see that the correct solution $c_j = \delta_{jk}$ 
is a minimizer of Eq.~\ref{eq:eq4}.

\begin{table*}[htbp]
  \begin{center}
    \caption{
      Energies (in Hartrees) of the SCF solutions
      afforded by CDIIS, LIST, and stability analysis 
      for various systems. The LIST data were taken from Ref.~\cite{LISTb}.
      The geometries are the same as those used in 
      Refs.~\cite{LISTi,LISTb,Garza2012}.
      }
    \label{tab:1}
    \begin{tabular*}{1\textwidth}{@{\extracolsep{\fill}}  c c c c c c c c }
      \toprule
      System & Level of Theory & $E_\text{CDIIS}$ & $E_\text{LIST}$ & $E_\text{Stable}$
      & $E_\text{Stable} - E_\text{CDIIS}$ & 
      $E_\text{Stable} - E_\text{LIST}$ & 
      $E_\text{LIST} - E_\text{CDIIS}$ \\
      \toprule
      SiH$_4$ & SVWN5/6-31G* & 
      $-290.45782$ & $-290.45770$ & $-290.45782$ & $0.0 \times 10^0$ & $-1.2 \times 10^{-4}$ & $-1.2 \times 10^{-4}$ \\
      $[$Cd(Im)$]^{2+}$ & B3LYP/3-21G &
      $-5667.00871$ & $-5667.00872$ & $-5667.00941$ & $-7.0 \times 10^{-4}$ & $-6.9 \times 10^{-4}$ & $-1.3 \times 10^{-5}$ \\
      Ru$_4$(CO) & B3LYP/LANL2DZ & 
      $-488.69727$ & $-488.71067$ & $-488.71765$ & $-2.0 \times 10^{-2}$ & $-7.0 \times 10^{-3}$ & $-1.3 \times 10^{-2}$ \\
      UF$_4$ & B3LYP/LANL2DZ &  
      $-451.21184$ & $-451.23119$ & $-451.24006$ & $-2.8 \times 10^{-2}$ & $-8.9 \times 10^{-3}$ & $-1.9 \times 10^{-2}$ \\
      \toprule 
    \end{tabular*}
  \end{center}
\end{table*}

In the original LISTb paper~\cite{LISTb}, 
the better convergence properties of LISTb, as compared to LISTd,
were attributed 
to an alleviation of the linear dependency problem of LISTd 
by the transposition. 
This assertion is not correct;
the ill-conditioning of the LISTd matrix $\mathbf{A}$ 
is determined by its condition number, which is 
the ratio between the largest and smallest singular values 
of $\mathbf{A}$. 
The singular value decomposition of $\mathbf{A}$ is 
$\mathbf{A} = U \Sigma V^T$, whereas for LISTb 
$\mathbf{A}^T =  V \Sigma U^T$. 
The transposition therefore does not alleviate the 
linear dependency of $\mathbf{A}$ in any way. 
An argument based on Cramer's rule 
was also given in Ref.~\cite{LISTb}, 
which pointed out that 
LISTd would tend to yield coefficients of large magnitude 
and opposite signs when close to convergence. 
As is common knowledge in numerical analysis, 
subtracting two large numbers of opposite signs is unwise 
because of the possibility of catastrophic cancellation.
However, this explanation is still unsatisfactory because the
magnitudes of the coefficients would need to be extremely 
large---comparable to the inverse of machine precision---and LISTd
performs poorly even when far from convergence~\cite{LISTi,LISTb}.
The argument for the improved acceleration of LISTb over LISTd
based on the minimization of approximate 
error functions seems therefore much more plausible
than the aforementioned explanations and simultaneously clarifies
why LISTi is also better than LISTd.


\emph{\textbf{Discussion.}}
The most widely utilized version of DIIS employs the
commutator $[F_{i}, D_{i}]$ as error vector
since $[F,D]=0$ is a necessary and sufficient
condition for an SCF solution \cite{Pulay2}.
This specific variant is commonly known as commutator-DIIS or CDIIS. 
We have seen here that LIST methods can be formulated as DIIS 
in the general framework outlined originally by Pulay~\cite{Pulay1}.
Considering this equivalence between LIST and DIIS, 
and how well-established the latter is, 
it is unsurprising that LIST can provide convergence 
acceleration. 
However, of the three different flavors of LIST, just two 
(LISTi and LISTb) work properly since only these
minimize a suitable error function.
More specifically, LISTi and LISTb minimize errors 
associated with necessary (albeit not sufficient) conditions 
for convergence, whereas LISTd minimizes a function
which is not related to necessary or sufficient conditions for 
convergence.

Based largely on the poor performance of CDIIS for the 
singlets of SiH$_4$ (with a broken bond), 
[Cd(Im)]$^{2+}$, Ru$_4$(CO) and UF$_4$, 
the authors in Refs.~\cite{LISTi,LISTb} concluded that LIST 
methods were superior to CDIIS.
For these systems, CDIIS appeared to be trapped in states higher 
in energy than the LIST solutions.
However, no stability analysis was carried out to verify 
whether the LIST solutions were high-energy states too.
Table \ref{tab:1} compares the energies reported for 
LIST~\cite{LISTb} with those from our calculations in 
\emph{Gaussian}~\cite{Gaussian} 
using CDIIS [\texttt{SCF=(CDIIS,NoDamp)} keyword] and 
RHF $\to$ RHF stability analysis [\texttt{Stable=(RRHF,Opt)}].
All our CDIIS calculations use the default iterative subspace of twenty
vectors, the Harris guess~\cite{Harris}, 
and tight convergence criteria.
Cartesian $d$ and higher functions
(\texttt{6D}, \texttt{10F}) were used to compare
the energies from \emph{Gaussian} with those of NWChem~\cite{Nwchem}
(the package used
in Refs.~\cite{LISTi,LISTb}) for a given basis.
We also employed the \texttt{Integral=Ultrafine} keyword 
and, based on calculations with different grid sizes, we estimate 
the error due to grid size in our comparisons with the data from 
Ref.~\cite{LISTb} to be about 0.1 mHartree.
Based on these considerations, it appears like LIST and CDIIS 
converge to the same solutions for SiH$_4$ and [Cd(Im)]$^{2+}$.
For Ru$_4$(CO) and UF$_4$, LIST converges to lower energy states 
than DIIS; however, the data obtained 
from stability analysis $E_\text{Stable}$
reveals that the LIST solutions
are high-energy states too.
We also carried out LIST calculations starting from the 
Harris guess---atomic densities were used as initial 
guess in Refs.~\cite{LISTi,LISTb}---with 
our own implementation of LIST
in \emph{Gaussian}; for SiH$_4$ and [Cd(Im)]$^{2+}$ these converge
to the same solutions as CDIIS; 
for Ru$_4$(CO) and UF$_4$, LIST converges to 
energies of $-488.71067$ and $-451.23012$ Hartrees, respectively. 
Hence, it seems definitive that 
LIST has the same problem as CDIIS for these 
systems---both converge to high-energy states---, 
in agreement with what would be expected from the 
equivalences between LIST and DIIS derived above. 
We also note that we did not find 
any convincing indication of LIST superiority in our previous 
numerical studies~\cite{Garza2012}.

It is also germane to point out that---for
SiH$_4$, Ru$_4$(CO) and UF$_4$---wavefunctions much lower in 
energy can be found if one allows the initial guess to 
break spin symmetry or via RHF $\to$ UHF stability analysis.
The symmetry breaking is an indication of static correlation.
The symmetry-adapted single Slater determinant approximation 
therefore breaks down, which is most likely the reason for 
the many unphysical solutions that can be found
in these systems.
Convergence to these high-energy states is a problem 
that neither LIST nor CDIIS can truly fix because this 
depends crucially on the initial guess.
Even using more robust (and expensive) algorithms, 
such as Bacskay's quadratically convergent SCF 
procedure~\cite{Bacskay1981}, we were not able to find the 
lower energy solutions obtained via stability analysis in 
Table~\ref{tab:1}.

Another problem shared by LIST and DIIS is that they may 
result in oscillations~\cite{LISTi,LISTb,Garza2012}.
This behavior arises because minimization of 
an approximate error function does not force convergence. 
A way to solve this issue is to use methods that ensure
a decrease in the energy at every iteration, as 
this guarantees convergence to a local 
minimum~\cite{Cances2000,LeBris}.
This is the motivation behind the optimal damping 
algorithm~\cite{Cances2000}
and its generalization in EDIIS~\cite{EDIIS}.
These techniques exploit the fact that, because of 
the \emph{aufbau} principle, all local energy minima 
for an idempotent density matrix are in the convex set 
$\tilde{\mathcal{P}}_{N}=\{\tilde{D}\in \mathcal{M}%
_{S}(N_{b}),\tilde{D}^{2}\leq \tilde{D},\text{Tr}(\tilde{D})=N\}$ 
($N_b$ is the number of basis functions; 
$\mathcal{M}_S (N_b)$ the set of square Hermitian matrices
of dimension $N_b$).
Thanks to this property, EDIIS can reduce the energy at every 
iteration without a significant increase in cost over DIIS
via an interpolation of previously iterated density 
matrices---the interpolation, rather than extrapolation,
is necessary to ensure $\tilde{D_{k}} \in \tilde{\mathcal{P}}_{N}$.
The interpolation has the side effect of making convergence 
slower as compared to DIIS, and thus EDIIS is often combined 
with DIIS to make the algorithm faster than the former and 
more robust than the latter~\cite{EDIIS,Garza2012}. 
Basically, EDIIS is used when far from convergence 
(as judged by the DIIS error) to bring the 
density matrix near the convergence region, where 
DIIS is most efficient.
Because of this improved robustness, a combination of 
EDIIS and DIIS has been used as the default option 
for SCF convergence in the \emph{Gaussian} suite of 
programs for many years~\cite{Gaussian}.


\textbf{\emph{Conclusions.}}
The LIST methods can all be formulated as extrapolation
techniques that minimize an approximate error function 
associated with a density matrix in the iterative subspace,
and thus
fall within the general scheme of DIIS first described by Pulay. 
This formulation also explains why LISTd has such a poor
performance---the DIIS error minimized is not a suitable one. 
The other LIST methods were derived from LISTd; however, 
they introduce approximations which lead 
to better error functions and thus have better convergence properties.
Nevertheless, because of the equivalences shown here,
they can hardly be better than DIIS and share the same problems as the 
commonly used CDIIS. 
From a formal perspective,
CDIIS appears to be more desirable than LIST 
as $[F,D]=0$ is a necessary and sufficient
condition for an SCF solution, whereas LISTi and LISTb
minimize error functions corresponding to necessary
(but not sufficient) conditions for convergence.
Furhtermore, 
any extrapolation technique that consists of solving a 
linear system of the form 
$\sum_j c_j b_{ij} = 0, \forall i$ restricted to 
$\sum c_j =1$, is in fact a form of DIIS.
Finally, we have also presented numerical data demonstrating that the 
alleged superiority of LIST over CDIIS is incorrect. 
Claims in the literature
that LIST outperforms CDIIS,
EDIIS, and their combinations are 
inaccurate. 
The mathematical proofs here presented make 
any assertion of inherent
LIST superiority unjustifiable. 

\textbf{\emph{Acknowledgments.}}
This work was supported by the National Science Foundation CHE-1110884 and
the Welch Foundation (C-0036). 
We thank Dr. Thomas Henderson for a critical reading of this 
manuscript.

\end{document}